\documentclass[usenatbib,usegraphicx]{mn2e}
\usepackage{aas_macros}
\usepackage{natbib}
\usepackage{amsmath,amssymb,amsfonts,bm}
\usepackage{subfigure}
\usepackage{longtable}
\usepackage{graphicx}
\usepackage{graphics}
\usepackage{keyval}
\usepackage{trig}
\usepackage{float}
\usepackage[dvipsnames]{color}
\usepackage{calc}
\usepackage{url}
\usepackage{hyperref}
\def\beq{\begin{equation}}
\def\eeq{\end{equation}}
\def\bey{\begin{eqnarray}}
\def\eey{\end{eqnarray}}
\def\Myr{\, {\rm Myr} }

\def\pc{\, {\rm pc} }

\def\mpc{\, {\rm Mpc} }
\def\msun{\rm M_\odot}
\def\sun{\odot}
\def\Msun{\rm M_\odot}

\def\lvsun{L_{\odot V}}
\def\mvsun{M_{\odot V}}
\def\kms{\, {\rm km \, s}^{-1} }
\def\vg{{\bf g}}

\def\a0{$a_0$}

\def\ge{{\bf g}_{\rm ext}}

\def\mvir{M_{\rm vir}}
\def\rvir{r_{\rm vir}}
\def\mgc{M_{\rm GC}}
\def\vc{v_{\rm c}}
\def\mb{M_{\rm b}}
\def\lv{L_{\rm V}}
\def\sn{S_{\rm N}}
\def\gn{g_{\rm N}}

\begin{document}
\title{The specific frequency and the globular cluster formation efficiency in Milgromian dynamics
}
\author[]{Xufen Wu$^{1}$, Pavel Kroupa$^{1,2}$ \\ 
$^{1}$ Argelander-Institut f\"{u}r Astronomie der Universit\"{a}t Bonn, Auf dem H\"{u}gel 71, D-53121 Bonn, Germany \\
$^{2}$ Helmhotz-Institit f\"{u}r Strahlen-und Kernphysik, Universit\"{a}t Bonn, Nussallee 14-16, D-53115 Bonn
}
\maketitle

\begin{abstract}
Previous studies of globular cluster (GC) systems show that there appears to be a
universal specific GC formation efficiency $\eta$ which relates the
total mass of GCs to the virial mass of host dark matter
halos, $\mvir$ \citep{Georgiev3,Spitler_Forbes2009}. 
In this paper,
the specific frequency, $\sn$, and
specific GC formation efficiency, $\eta$, are derived as functions of $\mvir$ in Milgromian dynamics, i.e., in modified Newtonian dynamics (MOND). 
In Milgromian dynamics, for the galaxies with GCs, the mass of the GC system,
$\mgc$, is a two-component function of $\mvir$ instead of a simple
linear relation. 
An observer in a Milgromian universe, who interprets this universe as being Newtonian/Einsteinian, will incorrectly infer a universal constant fraction between the mass of the GC system and a (false) dark matter halo of
the baryonic galaxy. 
In contrast to a universal constant of $\eta$, in a Milgromian universe,
for galaxies with $\mvir \le 10^{12}\msun$, $\eta$ decreases with the increase of $\mvir$, while for massive
galaxies with $\mvir>10^{12}\msun$, $\eta$ increases with the increase of $\mvir$.

\end{abstract}
\begin{keywords}
galaxies: star clusters - galaxies: general - galaxies: 
stellar content - galaxies: kinematics and dynamics
\end{keywords}
\section {Introduction}
The previous studies on the relation between the specific frequency,
$\sn$, of globular clusters (GCs), i.e., the number of GCs per unit luminosity of the host galaxy,
and the luminosity of the host galaxy, imply
that the formation of GCs has a universal mode,
irrespective of galaxy morphology \citep{Georgiev3}. Parameters such as the specific frequency of GCs are considered as
important tools to study star formation in galaxies and to study the
formation and evolution of the galaxies
\citep{Kissler-Patig2000,vandenBergh2000,Harris1991,Brodie_Strader2006}.
The observations of GCs show that $\sn$ varies greatly
with galaxy luminosity. In addition, there is an overall
trend of $\sn$ with galaxy luminosity: $\sn$ decreases with decreasing luminosity for the most massive Elliptical galaxies to reach a broad plateau at Milky Way class galaxies. In the dwarf galaxy regime $\sn$ bifurcates by being zero for star-forming galaxies while achieving very high values for some dwarf elliptical and dwarf spheroidal galaxies
\citep{Harris1991,Peng08,ML07,Spitler08,Georgiev3}.
\citet{McLaughlin1999} analyzes the total mass of the GC systems,
$\mgc$, and the mass of baryons, $\mb$, 
in three giant galaxies, and finds that these
two global observables have a constant ratio, $\mgc/\mb~=~0.26\%$, within any galactocentric radius inside each galaxy and for different galaxies. 
However,
\citet[][Fig. 2]{Spitler_Forbes2009} show that the relation $\mgc(\mb)$ has a different slope for massive and
dwarf galaxies, thus $\mgc(\mb)$ is not a linear function for
the whole mass range of the host galaxies. 

According to the Standard
Model of Cosmology, dark matter halos drive the formation and
the evolution of galaxies. Correlations between dark matter halos
and various baryonic properties of the galaxies have been
discovered. For example, the correlation between the specific frequency
of GCs and the virial mass of the hosting dark
matter halo has been interpreted to mean that more massive dark
matter halos yield a larger efficiency of GC formation.
It is suggested by
\citet{Blakeslee1997} and \citet{Blakeslee1999} that in both giant elliptical galaxies and also in brightest cluster galaxies (BCGs), the ratio between the total
mass of GCs in a galaxy and the total virial mass of the galaxy (with dark halo), $\mvir^{\rm tot}$, i.e., $\eta_{\rm tot}\equiv \mgc/\mvir^{\rm tot}$, is a more fundamental constant. \citet{Blakeslee1999} finds that $\eta_{\rm tot}~=~1.71\pm0.53 \times 10^{-4}$. 
\citet{Spitler_Forbes2009} study the total mass of GCs and the virial
mass of their hosting dark halo in a cold dark matter (CDM)
universe, and show that the total mass of GCs is proportional to the virial mass of a galaxy for galaxies with different morphologies,
except for a few local group dwarf galaxies, which are thought to be due to a sample bias in the stellar masses of these low mass galaxies. For such a relation, the GC formation effeciency $\eta_{\rm tot}=7.08\times10^{-5}$ can be derived. The recent work by \citet{Georgiev3} gives a mean value of
$\eta_{\rm tot}=5.5\times10^{-5}$ from the $S_L(M_V)$ relation for a wide range of galaxy masses for both early-type and late-type galaxies from a variety of observations \citep{ML07,Peng08,Spitler08,Georgiev_etal2008,Georgiev1,Georgiev2}, where $S_L$ is the specific GC luminosity \citep{Harris1991}.

\citet{Georgiev3} derive scaling relations of GCs, such as
$\sn$, specific mass and specific luminosity, as functions of dark matter halo
mass (including the baryonic matter in the halo) for both low mass and massive galaxies, which are, respectively, probably
regulated by supernova feedback and virial shock-heating of the
infalling gas. The halo masses of the galaxies are based on the cold dark matter (CDM) paradigm. However, in the derivation of \citet{Georgiev3} a 
universal central halo mass of the hosting
galaxies, $M_{0.3 {\rm kpc}} = 10^7\msun$, is used.
This value is extracted from the observations by
\citet{Strigari_etal2008}. Moreover, based on the co-added rotation curves of about 1000 spiral galaxies, it is found that $M_{0.3 {\rm kpc}}=1.35\times 10^7\msun$ \citep{Donato_etal2009}, which is consistent with \citet{Strigari_etal2008}'s observations.  
However, \citet{Kroupa_etal2010} show
that the values of the dark matter masses, $M_{0.3 {\rm kpc}}$, obtained from various
CDM cosmological simulations do not agree with
Strigari's observations. Therefore currently there are no existing
CDM halo models from cosmological simulation with such a central halo mass. Actually, the dark matter halo adopted in the devivation of \citet{Georgiev3} is an empirical combination of observations in the centre and of simulations in the outer regimes of a galaxy, rather than one that is naturally obtained from cosmological halo models.

On the other hand, \citet{Milgrom2009} and \citet{Gentile_etal2009} show
that there is an upper limit of the apparent central dark matter
surface density in Milgromian dynamics \citep{Milgrom1983,BM1984},
which is $\Sigma_M\equiv a_0/2\pi G=138 \msun \pc^{-2}$, and the
central (within $300\pc$) apparent halo projected mass is $1.24\times 10^7\msun$. This agrees
well with the observations by \citet{Strigari_etal2008} and
\citet{Donato_etal2009}. Here $a_0 \approx 3.7 \pc/\Myr^2$ is Milgrom's (1983) acceleration constant
\citep{Milgrom1983,BM1984,Milgrom2009,Famaey_McGaugh2012}. 
 In Milgromian dynamics each isolated baryonic mass of a galaxy, $M_b$, generates an effective force field which is mathematically equivalent in Newtonian dynamics to a logamithmic dark matter potential. 
Given the success of Milgromian dynamics on accounting for observed properties of galaxies in the central \citep{Milgrom2009} and outer regimes \citep[see the recent reviews of][]{Famaey_McGaugh2012,Kroupa_etal2012} and the problems of the $\Lambda$CDM model \citep{Kroupa2012}, we study here how the specific frequency of GCs and other quantities correlate with the phantom dark matter halos that surround each baryonic system.

The apparent (Newtonian) virial masses, $\mvir$, for galaxies used in \citet{Georgiev3}
and the apparent (Newtonian) dynamical mass-to-light ratios are studied in this paper in Milgromian dynamics. The relationship between $\mgc$ and $\mvir$ of galaxies is quantified. The GC scaling parameters, i.e., $\sn$ and $\eta$, are derived as functions of apparent CDM halo virial mass. We explicitly note that the following usage of language here: In a Milgromian universe there is no cold or warm dark matter which plays a dynamically significant role on galactic scales. Anyone who (wrongly) interprets the data by applying Newtonian dynamics will, however, (wrongly) introduce the presence of cold or warm dark matter in galaxies. We refer to this as apparent or phantom dark matter throughout this contribution.

\section{Analyses and results}
\begin{figure}{}
\begin{center}
\resizebox{9.cm}{!}{\includegraphics{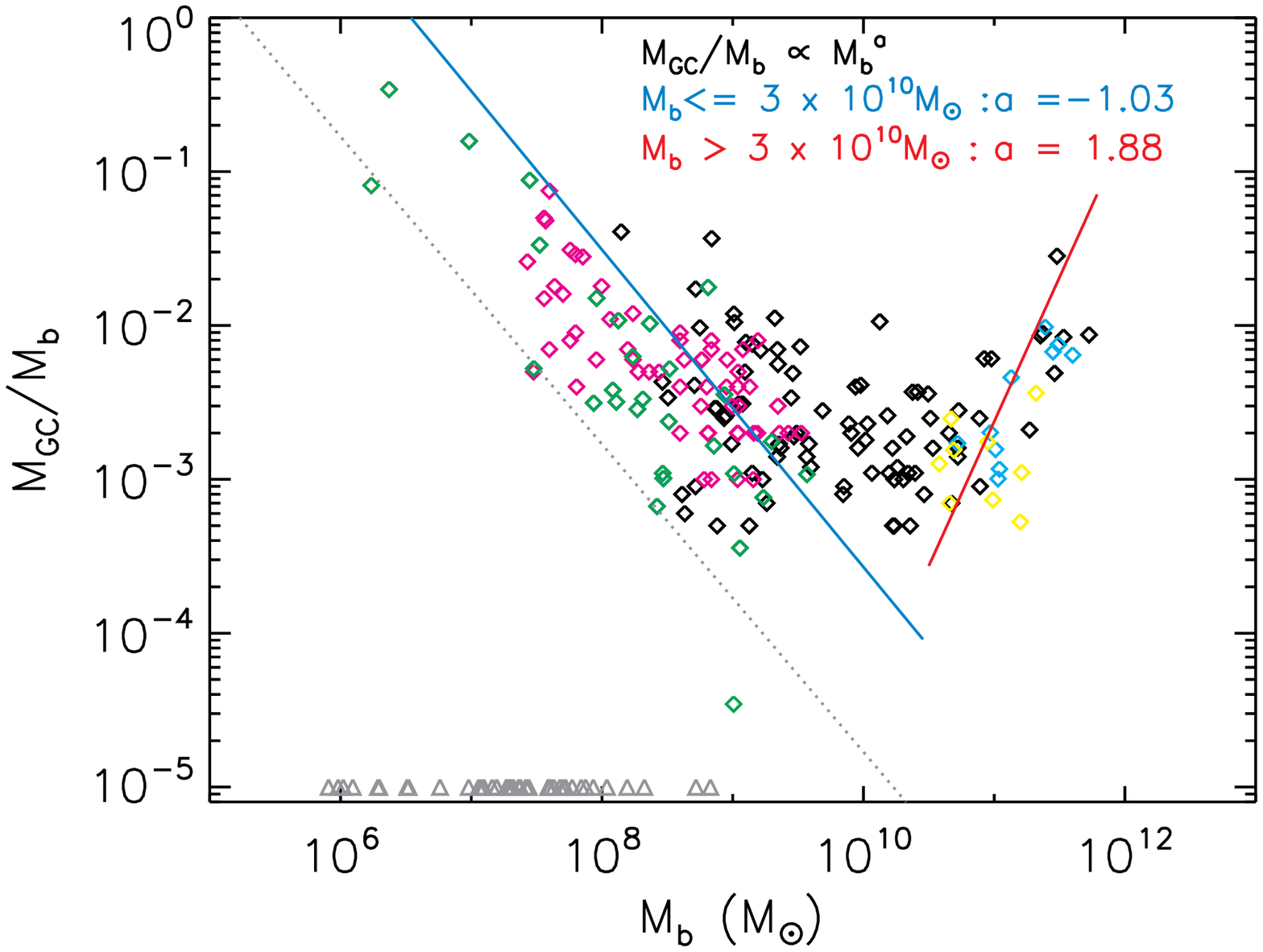}}
\makeatletter\def\@captype{figure}\makeatother \caption{The $\mgc/M_b$ ratio as a function of $M_b$.  The colours of the data points represent different sources of data: \citet[][black, for ellipticals]{Peng08}, \citet[][magenta, for dwarf ellipticals]{ML07}, \citet[][cyan, for ellipticals]{Spitler08}, \citet[][yellow, for sprials]{Spitler08}, \citet[][green, for nearby dwarf galaxies]{Georgiev1,Georgiev2}. The gray triangles at the bottom of the figure show galaxies without any GCs observed, while the dotted line shows galaxies only containing one GC. The blue and red lines show, respectively, the linear fit for the data points corresponding to $M_b \le 3 \times 10^{10}\msun$ and $M_b > 3\times 10^{10}\msun$, the least square fitting parameters are labelled in the figure.
}\label{mgcmb}
\end{center}
\end{figure}

The observed samples of galaxies are the same as those used
in \citet{Georgiev3}. A plot of $\mgc/M_b$ as a function of $M_b$ is shown in Figure \ref{mgcmb}. The $\mgc/M_b$ versus $M_b$ data is described as a three-part function of $M_b$. For galaxies with $M_b < 3\times10^{10}\msun$, $\mgc/M_b$ either decreases as $M_b$ increases, or $\mgc/M_b=0$, i.e., in some dwarf galaxies no GC systems are observed. For galaxies with $M_b>3\times 10^{10}\msun$, $\mgc/M_b$ increases as $M_b$ increases. For the galaxies with GCs, the fit parameter for the relation of $\mgc/M_b$ and $M_b$ are given in the figure.

\subsection{The apparent virial mass and apparent dynamical mass-to-light ratio}\label{secmvir}
Milgrom's original proposal for a new effective gravitational dynamics \citep[][]{Milgrom1983,BM1984} which is sourced purely by baryonic matter, has passed all tests that have been performed until now on galactic scales. It accounts for the
baryonic Tully-Fisher relation, the shapes of rotation curves of
galaxies, the dark matter effect in tidal dwarf galaxies, the
universal scale of baryons, and the projected surface density of
apparent dark matter within the core radius of the apparent dark
matter halos
\citep{TF1977,Milgrom_Sanders2003,Sanders_Noordermeer2007,
Gentile_etal2007,Gentile_etal2009}. In Milgromian dynamics, the
Newtonian gravitational acceleration $\gn$ is replaced with
$g=\sqrt{\gn a_0}$ when the gravitational acceleration is much
smaller than $a_0 \approx c\Lambda^{1/2}$, while the strong gravity
behaves Newtonian. Here $\Lambda$ is the cosmological constant. The
weak field approach empirically links the gravity in galaxies with
the baryonic distribution without any cold or warm dark matter. The
modified Poisson equation in Milgromian dynamics is 
\beq \nabla
\cdot [\mu(\frac{|\vg|}{a_0})\vg]=-4\pi G \rho_b,
\eeq 
where $\rho_b$
is the baryonic density and $\mu(\frac{|\vg|}{a_0})$ is an
interpolating function. $\mu \rightarrow |\vg|/a_0$ when $g\ll a_0$
and $\mu \rightarrow 1$ when $g\gg a_0$ \citep{BM1984}.
\citet{Famaey_McGaugh2012} review the observational sucesses and
problems of Milgrom's dynamics. Milgromian dynamics can be related to
space-time scale invariance \citep{Milgrom2012a,Kroupa_etal2012} and quantum-mechanical processes in the vacuum (\citealt{Milgrom1999,Zhao2008}, see also Appendix A in \citealt{Kroupa_etal2010}). 

From the weak field approach, the circular
velocity at large radii of a galaxy follows as being
\beq\label{eq_vc}\vc=(Ga_0\mb)^{1/4}.\eeq
This implies flat rotation curves and thus, in terms of Newtonian dynamics, an apparent phantom dark matter
distribution with an isothermal profile at large radii. Here $G$ is
the gravitational constant and $\mb$ is the total baryonic mass of a
galaxy.
The apparent virial radius and the mass of the phantom dark matter halo
can be derived from the asymptotic behaviour of the rotation curve. 
The apparent phantom dark matter halo (Newtonian)
virial masses, $\mvir$, and the apparent (Newtonian) dynamical
mass-to-light ratios of the \citet{Georgiev3} galaxies are here studied in
Milgromian dynamics. 
The virial radius is defined as the radius where the average density of
enclosed apparent dynamical matter which leads to the flat rotation curve of a galaxy (baryonic matter and phantom dark matter) is 200 times the critical density, $\rho_{\rm crit}$, of the universe, i.e.,
\beq
\mvir\equiv p\rvir^3  =\vc^2\rvir/G.
\eeq
The virial mass of the phantom
dark matter halo can be written as  
\beq\label{eq_mvir}\mvir=\vc^3p^{-1/2}G^{-3/2},\eeq
where $p=\frac{4}{3}\pi \times 200\rho_{\rm crit}$, here $\rho_{\rm crit}=\frac{3H^2}{8\pi G}$ is the critical density of the universe and $H$ is the Hubble
constant, $H=70\kms\mpc^{-1}$. Therefore the
apparent virial mass is proportional to $\mb^{3/4}$. The apparent
virial mass is plotted in dependence of the V-band absolute  magnitude, $M_V$, of the galaxies used in
\citet{Georgiev3} in the upper left panel of Figure
\ref{mvir}. Clearly, there is a tight anti-correlation between
$\mvir$ and $M_V$. Such a correlation can indeed be expected, since the apparent virial mass is determined by the total mass of the baryonic matter in Milgromian dynamics.  

The observed $M_V$ and $L_V$ data of the galaxies (here $L_V/\lvsun=10^{-0.4(M_V-\mvsun)}$, where $\lvsun$ and $\mvsun$ are the luminosity and absolute magnitude of the Sun in the V band) and of $\mgc$ in each galaxy used in this paper are originally taken from various observations \citep{ML07,Peng08,Spitler08,Georgiev_etal2008,Georgiev1,Georgiev2}, and the data for $M_b$ of the galaxies are obtained by performing the same calculation as \citet[][see \S 3.2.1 in their paper]{Georgiev3}: The masses of most of their sample galaxies are computed by using a luminosity-$M/L$ relation derived from \citet{Bell_etal2003}, except for the dwarf galaxies from \citet{ML07}, for which a constant mass to light ratio of $3$ is assumed. Combining Equations \ref{eq_vc} and \ref{eq_mvir}, the virial mass of the phantom dark matter halo is derived as
\beq\label{eq_mvir2} \mvir=(Ga_0\mb)^{3/4}p^{-1/2}G^{-3/2}.\eeq The apparent dynamical mass-to-light (V band) ratios of
the galaxies, $\mvir/\lv$, are shown in the lower left panel of Figure \ref{mvir}.
$\mvir/\lv$ correlates with $M_V$, for
brighter galaxies the $\mvir/\lv$ ratios are small, while for
fainter galaxies the $\mvir/\lv$ ratios are large.

\subsection{$\mgc$, Specific frequency $\sn$ and $\eta$ values}\label{secmass}
\begin{figure}{}
\begin{center}
\resizebox{9.cm}{!}{\includegraphics{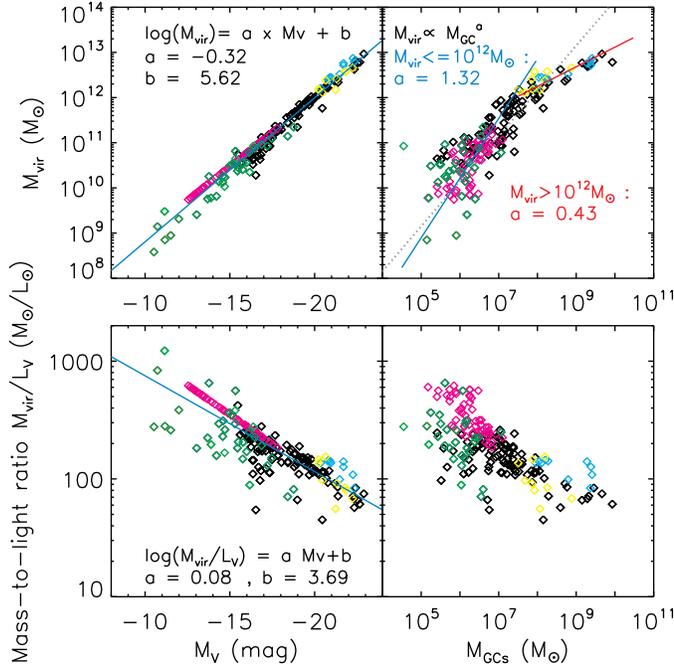}}
\makeatletter\def\@captype{figure}\makeatother \caption{The
apparent phantom cold or warm dark matter virial masses versus absolute magnitude, $M_V$, of
galaxies (upper left panel), apparent Newtonian dynamical mass-to-light
ratios in Milgromian dynamics versus $M_V$ (lower left). The blue lines on left panels show
the linear fit to the data points, and the least square fitting parameters are labelled 
in the left panels.
The upper right panel shows the
apparent virial masses against the total GC masses. The blue and red lines show, respectively, the linear fit for the data points corresponding to $\mvir \le 10^{12}\msun$ and $\mvir>10^{12}\msun$, and
the dotted
line shows the linear fit for CDM halos from
\citet{Spitler_Forbes2009}. The lower right panel shows the
apparent Newtonian dynamical mass-to-light ratios in Milgromian dynamics against the total GC 
masses. The colours of the data points are the same as in Figure \ref{mgcmb}.}\label{mvir}
\end{center}
\end{figure}

It has been suggested that $\mgc \propto \mvir$
in CDM models of galaxies \citep{Blakeslee1997,Blakeslee1999,Spitler_Forbes2009}. $\mgc$ and apparent $\mvir$ in Milgromian dynamics are compared in the upper
right panel of Figure \ref{mvir} ($\log$ is $\log_{10}$ hereafter). The slopes of $\log \mvir$ in Milgromian dynamics and $\log \mgc$ are given in this panel by a linear fit function, for $\mvir\le 10^{12}\msun$ on the top left and for $\mvir > 10^{12}\msun$ on the bottom right. For galaxies with
apparent $\mvir \le 10^{12} \Msun$, $\mvir \propto \mgc^{4/3}$, while
for massive galaxies with apparent $\mvir > 10^{12}\msun$, $\mvir
\propto \mgc^{0.43}\approx \mgc^{3/7}$. However a rather good agreement with the empirical CDM scaling (\citealt{Spitler_Forbes2009}, see the gray line in the upper right panel of Fig. \ref{mvir}) is evident.
For the massive galaxies, in Milgromian dynamics, the ratio
$\mgc/\mvir$ is larger than for the less massive galaxies, since
$\mvir/\mb$ is smaller for the massive galaxies. 
The above trend comes from the observed non-universality of $\mgc/\mb$ (see Figure \ref{mgcmb}) and the one-to-one relation between $\mvir$ and $M_b$.
The $\mvir/\lv$ values of the galaxies are plotted in dependence of $\mgc$ in the lower right
panel of Figure \ref{mvir}. $\mvir/\lv$ decreases
faster when $\mgc$ is increasing and $\mgc$ is smaller than
about $2\times 10^7\msun$ (i.e., where $\mvir\le 10^{12}\msun$),
and the decreasing trend becomes shallower for
systems with larger $\mgc$ (i.e., where the $\mvir>10^{12}\msun$).
This agrees with the trend of $\mvir$ as a function of $\mgc$.

\begin{figure}{}
\begin{center}
\resizebox{9.cm}{!}{\includegraphics{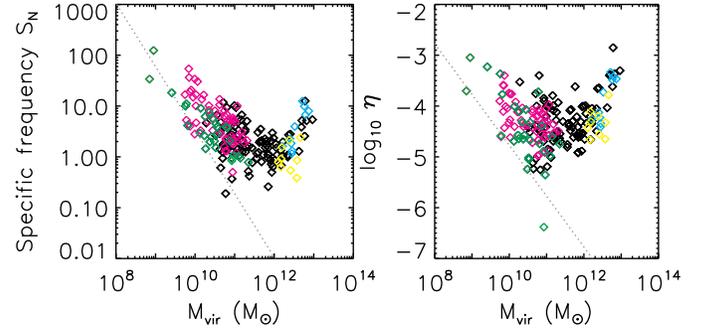}}
\makeatletter\def\@captype{figure}\makeatother \caption{Left panel:
the specific frequency of GCs as a function of
apparent virial mass for galaxies from \citet{Georgiev3}. Right
panel: the ratios of $\mgc/\mvir$ as a function of $\mvir$. The symbols are defined as in Figure \ref{mvir}. The dotted gray lines in both panels show the $\sn$ and $\eta$ values for galaxies hosting only one GC, and the areas below the gray dotted lines correspond to galaxies hosting fewer than one GC.
}\label{sn}
\end{center}
\end{figure}

From \S \ref{secmvir},  $\mvir \propto \mb^{3/4}$ in Milgromian dynamics. It is known that $\sn =N_{\rm GC}\times 10^{0.4(M_V+15)} \propto \mgc \lv^{-1}$. Since the stellar mass to light ratios of galaxies, $M_b/L_V$, are determined by the stellar population and stellar evolution, and also the dark baryonic components like hot gas, the relation of $M_b$ and $L_V$ is not simply $M_b\propto L_V$. A power law relation between $L_V$ and total baryonic mass in a galaxy is fitted in \citet{Georgiev3}, which is $M_b\propto L_V^{1.11} \approx L_V^{10/9}$. So $\sn \propto \mgc \mvir^{-6/5}$.
Therefore $\sn$ is a function of apparent $\mvir$:
\bey\label{snmvir}
\sn &\propto& \mvir^{3/4} \times \mvir^{-6/5}=\mvir^{-9/20},~~~~\mvir \le 10^{12}\msun \nonumber\\
\sn &\propto&
\mvir^{7/3}\times\mvir^{-6/5}=\mvir^{17/15},~~~~\mvir>10^{12}\msun \eey The specific frequency $\sn$ 
is shown in dependence of $\mvir$ 
in the left panel of Figure \ref{sn}. The values of $\sn$ are from
\citet{Georgiev3}. It is confirmed that the trend of data points has two
components, as expected in Eq. \ref{snmvir}. The two-component
function for galaxies with GCs implies that for low mass galaxies with
$\mvir \le 10^{12}\msun$ the number of GCs decreases with the apparent virial
mass; for massive galaxies with $\mvir>10^{12}\msun$, the
number of GCs increases with the apparent halo mass.

Since apparent (phantom) dark matter halos in Milgromian dynamics have
different density profiles compared to the dark halos obtained
from CDM cosmological simulations, it is therefore unnecessary that the $\mgc/\mvir$ ratio is a universal constant. From the trend of $\mgc$ as a function of
$\mvir$ in the upper right panel of Figure \ref{mvir},
the trend of the $\eta$ values for galaxies with GCs can be obtained: for galaxies with apparent halo
mass $\mvir \le 10^{12}\msun$, $\eta \propto \mvir^{-1/4}$; for
massive galaxies with  $\mvir >10^{12}\msun$, $\eta \propto \mvir^{4/3}$.
Thus the mass fraction of GCs in galaxies decreases with increasing apparent 
halo virial mass
when the galaxies are less massive than $10^{12}\msun$, and for the
massive galaxies the mass fraction of GCs increases with increasing 
apparent virial mass. 

The $\eta$ values for Georgiev's galaxies
\citep{Georgiev3} are shown in the right panel of Figure \ref{sn}, and they
argee with what is expected from the above discussion.
The function of $\eta$ in
Milgromian dynamics comes from the $100\%$ conspiracy of apparent dark matter
with baryons, and implies non-constant $\mgc/\mb$ ratios for the
GC-hosting galaxy systems. This is a natural expectation if the star formation rate (SFR) of a galaxy and thus its production of massive clusters \citep{Weidner_etal2004} depends on the depth of the potential well, although details need to be worked out.

\section{Discussion and summary}
In this work, it has been found that the apparent virial mass of the phantom dark matter halo (observed with Newtonian eyes) predicted in Milgromian dynamics
tightly anti-correlates with the V-band galaxy absolute magnitude
$M_V$, and that the apparent Newtonian dynamical mass-to-light ratios
correlate with $M_V$ in a Milgromian dynamics universe. 
The relationship of total GC mass in a galaxy
and the apparent dark matter halo virial mass of the hosting galaxy in Milgromian
dynamics has been studied here. It follows that $\mgc$ is a two-part
function of the apparent dark matter halo virial mass $\mvir$ for galaxies with GCs. 
For dwarf galaxies with $\mvir \le 10^{12}\msun$,
$\mgc \propto \mvir^{3/4}$, while for massive galaxies with
$\mvir>10^{12}\msun$, $\mgc \propto \mvir^{7/3}$. Therefore, the overall
mass of a GC system increases more slowly with the increase of
apparent virial mass for galaxies with shallower potential wells (i.e., $\mvir\le 10^{12}\msun$),
whereas the total mass of a GC system increases more rapidly with the
increase of the apparent virial mass for galaxies with deeper
potentials (i.e., $\mvir > 10^{12}\msun$). 

For galaxies with GCs, the specific GC formation
efficiency is $\eta \propto \mvir^{-1/4}$ for galaxies with
$\mvir\le 10^{12}\msun$, while for massive galaxies with $\mvir
> 10^{12}\msun$, $\eta \propto \mvir^{4/3}$, in contrast to a universal constant
$\eta$ obtained assuming CDM halos to be real.
The scaling relations of specific frequency $\sn$ and specific GC
formation efficiency $\eta$ as functions of apparent halo mass are different
to what is derived from CDM models. The two-component functions of
$\sn(\mvir)$ and $\eta(\mvir)$ indicate that GCs form more efficiently in
massive galaxies in Milgromian dynamics and in dE galaxies with small apparent
virial mass.

\section{Acknowledgments}
Xufen Wu gratefully acknowledges support through the Alexander von
Humboldt Foundation. We thank Iskren Georgiev for providing the observed 
data from Georgiev et al. (2010) and for the useful comments to the manuscript. 


\bibliographystyle{mn2e}
\bibliography{sn}
\end{document}